\documentclass[11pt]{article}

\usepackage{epsfig} 
\usepackage{graphicx} 
 
\begin{document} 
\thispagestyle{empty}     
 
\begin{flushright} 
UOSTP-03-104\phantom{bcdi}\\ 
{\tt hep-th/0308027}\phantom{ba} 
\end{flushright} 
\vspace{.3cm} 
 
\renewcommand{\thefootnote}{\fnsymbol{footnote}} 
\centerline{\Large \bf  
Holography with Timelike Bulk Hypersurfaces 
} 
 
\vskip 1.5cm   
\centerline{ \large Dongsu Bak\footnote{email: dsbak@mach.uos.ac.kr} 
} 
 
\vskip 10mm    
\renewcommand{\thefootnote}{\arabic{footnote}} 
\setcounter{footnote}{0} 
 
\centerline{\it 
Physics Department, University of Seoul, Seoul 130-743, Korea} 
\vskip 3mm     
\vspace*{0.4cm} 
\vskip 30mm 
 
\baselineskip 18pt 
 
\begin{quote} 
We propose a  new version of  holographic principle.  
This proposal extends the holographic principle based on the 
lightsheet to the one 
constraining 
the entropy passing through 
 bulk hypersurface  of timelike geodesics  
by the boundary area divided by $4G$. 
We give a proof of the proposal in the classical regime  
 based on a simple local entropy 
condition. 
\end{quote} 
\vskip 2cm 
\centerline{\today} 
\newpage 
\baselineskip 18pt 
 
\def\nn{\nonumber} 

\section{Introduction}

The holographic 
principle may be a key paradigm for the understanding of  
the true nature of   gravity and string theories.   
As  suggested  first by 't Hooft and Susskind, it  states 
that the degrees of freedom for the gravitational system 
  reside not in the bulk   
but in its boundaries\cite{TH,SU}.  
Furthermore the number of boundary degrees of  
freedom per Planck area is constrained not to exceed  
unity.  This principle clearly contradicts with the current framework of  
quantum  
field theories  because their numbers of  
 degrees 
scale like the bulk volume not by the boundary area.  
These ideas are precisely exemplified in  
the formulation of AdS/CFT correspondence\cite{MA,SW}. 
When the gravity and field theories coexist like our Universe, 
the UV structure of 
 quantum field theories must 
 be modified 
drastically  in order to have fundamental degrees limited by their  
boundary areas.

Application of  
 the above holographic principle to more general spacetime  
including cosmology 
is problematic\cite{FS}. For the flat FRW universe as an  
example, the matter entropy  scales like  
the coordinate volume and 
the  bulk entropy  eventually  wins over the area bound 
for large enough region.  
The authors of \cite{FS} suggested that 
the entropy crossing  past lightlike region generated from the  
particle horizon should be bounded by the area of the  
particle horizon. Based on the subsequent works\cite{BA}, 
Bousso has given a covariant version of the holographic  
principle 
that may be applicable for general backgrounds\cite{BO}. 
It   states that the entropy on the 
lightsheet $L$ is  bounded by its boundary area divided by $4G$.  
 The lightsheet  
is  defined by the bundle 
of lightlike geodesics  that are orthogonally generated from the 
boundary surface with nonpositive expansion away from the boundary  
surface. The shape of the 
 boundary surface may be arbitrary once it is consistent with 
the nonpositive  expansion condition. 
 
More stronger bound is proposed by Flanagan, Marolf and Wald (FMW)\cite{FMW},  
asserting that 
 the entropy flux  
passing through  
 the lightsheet orthogonally  
generated from one 
boundary  and ending on another boundary  
is bounded by 
the difference of the boundary areas  divided by $4G$. 
Interestingly this version can be proven in the  
semiclassical limit 
provided  certain local conditions on the entropy are assumed\cite{BFM,ST}.  
The  
proof is utilizing  focusing nature of the lightlike geodesic 
based on the Raychaudhuri equation.

In this note,  
we   observe that the Raychaudhuri equation is also for the timelike  
geodesics, which plays an important role in the investigation of the global 
structures of spacetime including horizons and singularities. 
We extend  the above 
holography 
to the case where  the bulk hypersurface consists of  timelike  
geodesics.  
The proposal dictates that  
the entropy passing through the bulk hypersurface 
consisting of timelike geodesics 
orthogonally  
generated from a codimension two spacelike  
hypersurface $B$ and terminating on $B'$, 
should be bounded by 
 the difference of the boundary area divided  
by $4G$. The proof shall be given in the classical regime  
with simple well motivated 
local entropy conditions. 
 
There is a strong supportive argument for the above timelike  
holography. Consider a $d+1$ dimensional geometry of the form, 
$M \times S^1$, where $M$ is a $d$ dimensional spacetime and  
$S^1$ denotes a circle. Lightlike geodesics in the $d+1$ dimensions  
 correspond to  timelike geodesics in  the spacetime $M$ 
if the momentum along the circle is nonvanishing. 
This way the lightlike holography in $M \times S^1$ may lead to the 
timelike holography in the lower dimensional  
spacetime\footnote{We like to thank Dominic Brecher for pointing out this  
argument.  
There are also discussions about the dimensional reduction  
and lightlike holography in Refs. \cite{Horava,Brecher}.}.

This paper is organized as follows. In Section 2, we review the covariant  
holography proposed by Bousso and extended by FMW. In Section 3,  
we give the proof (in the classical regime) based on the simple  
local conditions on the entropy density. Section 4 deals with our  
proposal on the  
holography with timelike bulk hypersurface. 
We discuss a  proof of the statement starting from appropriate  
local conditions on the entropy flux. 
Last section is devoted for the discussions.


\section{Review of the lightlike holographic principle}  
For later use, we will review the basic construct of the  
holographic  
principle 
introduced by Bousso and later  
extended by FMW. 
 
The Bousso's holographic principle compares the entropy on the 
lightsheet $L$ with its boundary area divided by $4G$.  
The lightsheet  
is defined by the bundle 
of lightlike geodesics  that are orthogonally generated from the 
boundary surface with nonpositive expansion away from   
a spacelike
surface. The principle then dictates that  
\begin{equation} 
S_L \ \le \ {A\over 4G} 
\label{one} 
\end{equation} 
where $S_L$ is the entropy flux passing through the lightsheet 
and $A$ the boundary area.    
 
The lightsheet consists of  collections of lightlike geodesics  
$k^a=\left({\partial\over \partial \lambda}\right)^a$ emanated  
orthogonally 
from a given codimension-two  
spacelike surface $B$ where $\lambda$ is the affine parameter. 
The expansion,  
\begin{equation} 
\theta= \nabla_a k^a\,, 
\label{two} 
\end{equation} 
describes the changing rate of the logarithm of the cross sectional 
area of the light rays with respect to $\lambda $. 
The definition of  lightsheet requires the expansion  
 away from the boundary surface 
 to be  
 nonpositive.  The lightsheet should be terminated 
if the expansion turns into a positive value or the geodesics  
hit  singularities.  
 
Though the local concept of  entropy is not well defined, we introduce 
the entropy current density $s^a$, which may have a well defined 
physical meaning in the thermodynamic limit. The entropy flux density 
$s$ passing through the lightsheet is  given by 
\begin{equation} 
s= n_a s^a\,, 
\label{three} 
\end{equation} 
with the vector $n_a= -k_a\ (+k_a) $ normal to the future (past)  
directed  
lightsheet.  
The total entropy passing through the lightsheet is the volume  
integral, 
\begin{equation} 
S_L(B)= \int dy^{D-2}\sqrt{h(y)}\int d\lambda\, {\cal A}\, \, s\,, 
\label{four} 
\end{equation} 
where $\vec{y}$ is  
the coordinate for the boundary  and $h(y)$ is the  
determinant of the induced metric on the boundary.  
The area factor  
${\cal A}$ represents the growth of the unit area along 
the affine parameter  
\begin{equation} 
{\cal A}= exp\left[{\int^\lambda_0  d\tilde{\lambda}\,\,\,  
\theta(\tilde{\lambda})}\right]\,, 
\label{five} 
\end{equation} 
and the cross sectional area at $\lambda$ is then given by 
\begin{equation} 
 A(\lambda)=\int dy^{D-2}\sqrt{h(y)}\,\, {\cal A}(\lambda)\,. 
\end{equation} 
 
More stronger bound,  
\begin{equation} 
S_{L(B-B')} \ \le\  {1\over 4G}(A-A')\, \,, 
\label{four1} 
\end{equation} 
has been proposed by  
FMW, 
where $S_{L(B-B')}$ corresponds to the entropy flux  
passing through  
 the lightlike surface orthogonally  
generated from 
$B$ and ending on $B'$ before reaching caustic or  
singularities. One  compares this entropy to 
the difference of the area of $B$ and $B'$. This version of the  
entropy bound is the one we are interested in this note.

\section{Conditions for the holography} 
  
The latter version of the holography can be shown to hold 
in the classical regime with certain conditions on the entropy flux density 
are satisfied.  
 
Here we review the conditions and the proof given recently in Refs. 
\cite{BFM,ST}. The entropy conditions used are: 
$$  
{\rm i.}\ \ \,\,\,  s' \  \le \   2\pi T_{ab} k^a k^b  
$$ 
$$ 
{\rm ii.}\ \  s(0) \ \le \   -{1\over 4G}\theta(0) 
$$ 
The second condition is on the initial condition for the  entropy 
and the first corresponds to the local version of 
the Bekenstein bound as discussed  in Ref. \cite{BFM,BE,BO2}.

Based on these two conditions, the proof of the entropy bound is very  
simple in the semiclassical limit where one may use the Einstein equation, 
\begin{equation} 
R_{ab}-{1\over 2} g_{ab} R= 8\pi G\,  T_{ab}\, \,. 
\label{five1} 
\end{equation}

 The expansion of the bundle of the lightlike  
geodesics obeys 
the Raychaudhuri equation 
\begin{equation} 
\theta'= -{1\over D-2}\theta^2 - \sigma_{ab}\sigma^{ab}+ 
\omega_{ab}\omega^{ab}  - R_{ab} k^a k^b\, \,, 
\label{six} 
\end{equation} 
where $\sigma_{ab}$ and  $\omega_{ab}$ are the shear and the twist. 
Since the lightsheets are hypersurface orthogonal, the twist $\omega_{ab}$ 
vanishes. 
 
The FMW bound has a differential form.  
We note that the local inequality, 
\begin{equation} 
s   \  \le \   -{1\over 4 G} \theta        \,, 
\label{seven} 
\end{equation} 
implies the integral version (\ref{four1}). 
In particular considering $B'$  
infinitesimally  
away from $B$ in terms of the affine parameter, one gets the  
condition (ii) from the integral version. 
It is clear that the initial condition on the 
entropy should hold without any disposal because the part of  
 statement of 
the holography is the initial condition itself.  
Since $s(0)$ is not to do with how the boundary surface is curved,  
the initial condition is 
not satisfied for an arbitrary boundary surface.  
Rather at a given  
initial spacetime point, one  needs more curved boundary in order to 
have the negative of expansion larger than the initial entropy flux  
density. 
Therefore the initial condition puts restriction not on the initial  
entropy density but on the boundary surface. 
In Ref. \cite{BFM}, the authors insisted $\theta (0) \ \le \ 0$, which is 
clearly different from ours.

From the two conditions, the proof of the inequality (\ref{seven}) 
is quite simple.  Using (\ref{six}), the first condition  and the  
Einstein Equation, we have 
\begin{equation} 
-{1\over 4G} \theta' = {1\over 4G(D-2)}\theta^2  
+{1\over 4G} \sigma_{ab}\sigma^{ab}+ 
2\pi  T_{ab} k^a k^b 
\ \ge\    s'      \,. 
\label{eight} 
\end{equation} 
Hence together with the initial condition, one has 
\begin{equation} 
s=s_0+\int_0^\lambda d\tilde\lambda\,\,  s'(\tilde\lambda)\   
\le\  -{1\over 4G} 
\left(\theta_0 + \int_0^\lambda d\tilde\lambda\,\,  
\theta' (\tilde\lambda)\right)=  -{1\over 4G}\theta      \,. 
\label{nine} 
\end{equation}

A few comments are in order. The first  condition on the  
local change of 
the entropy is closely related to the Bekenstein bound as discussed  
recently in Ref. \cite{BFM,BE,BO2}. We shall not repeat the arguments here. 
One might try using  the following weaker condition; 
\begin{equation} 
s'\ \,\, \le\   2\pi T_{ab}k^a k^b + {1\over 4G(D-2)}\theta^2\ ? 
\label{ten} 
\end{equation} 
But the expansion is not such a physical quantity that is fixed 
solely by the spacetime point.  Rather it is determined 
by the point as well as by the shape of  
the  initial surface. On the other hand, 
$s'$ and $T_{ab}k^a k^b$ are the physical quantities (of course  
depending on the geodesic $k^a$) that are nothing to do with 
the shape of the initial surface. In this respect we view that the  
relation (\ref{ten}) is not desirable.

The saturation of the above holographic principle  
requires $\theta=0$ as well as $s'= 2\pi T_{ab}k^a k^b$.  
The $\theta=0$ condition is particularly strict. For example 
in the case of AdS   geometries, we do not have holographic screen  
that saturates the holographic bound. The reason is because   
the $\theta=0$ condition cannot hold along the lightsheet. 
Let us  take a  lightsheet in the IR region of AdS space, for which the  
spacelike projection theorem is applicable. Then the lightsheet 
records  all the entropy of the bulk inside.  
As discussed in detail in Ref. \cite{SW},  
the boundary area of AdS$_5\times S^5$ 
divided by $4G$ located at the IR region of AdS space 
agrees with the degrees of freedom count of $N=4$ SYM theory up to  
order one numerical constant. However, if the current formulation 
of the holographic principle 
is correct, the numerical coefficient cannot be unity because 
 the bound cannot be saturated. Namely  
$S_{Bulk}= S_{CFT} \ \le \   S_L  \ <\  {1\over 4G} A$ with $S_{Bulk}$ and  
$S_{CFT}$ being respectively the entropies of the bulk and the  
boundary $N=4$ SYM theory.    
Since the original conjecture is the equivalence between   
gravity with 
$N=4$ SYM theory in the  large N limit,  the non  
saturation is not a problem.

\begin{figure}[ht!] 
\centering \epsfysize=9cm 
\includegraphics[scale=0.7]{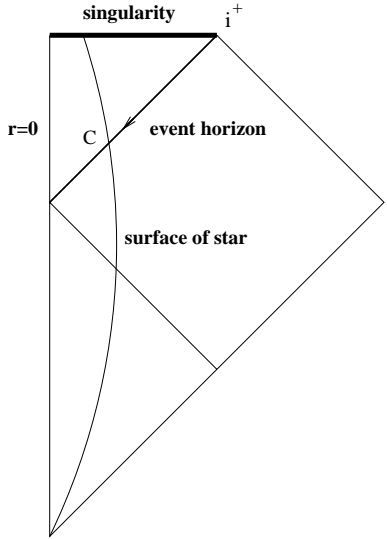} 
\caption{\small  We depict  the Penrose diagram of the spacetime 
representing the formation of the black hole by the gravitational collapse 
of a star. }  
\end{figure}

Let us give here examples where the bound is saturated. A rather trivial  
example is the case of flat boundary in the flat spacetime.  
The expansion  
and the entropy flux density along the lightsheet 
remain zero. The bound is saturated because the both sides are  
zero. 
Another is the spacetime of 
 a stationary black hole, where we take  
the event horizon  
 as lightsheet.  
There the expansion stays zero.  Hence the difference of the boundary areas 
is again  zero. Also  $s=0$ because  
nothing crosses the horizon due to the stationary nature of the black hole. 
Of course, we ignore the possible quantum corrections including the  
effect of Hawking radiation.

Finally we illustrate the use of holography for the spacetime    
 representing 
 the formation of black hole by the gravitational collapse of star. 
The relevant Penrose diagram is depicted in Fig. 1.  
Ignoring any quantum effect, the outside of the surface of the star 
is same as the vacuum Schwarzschild black hole solution with the  
horizon radius, say, $R_S$.  
The event horizon 
(the line with arrow) begins with $r=0$,  its radius grows 
toward the future infinity $i^+$ and reaches the stationary value 
$R_S$ once  the surface of star is crossed. The increase is  
monotonic in accordance with the theorem of horizon area for the matter 
content of the star 
satisfying the weak energy condition.   
Let us consider  the lightsheet starting from somewhere between $i^+$ and  
$C$ and ending up at $r=0$. (Thus $\lambda=0$ at the starting point.)  
Expansion is nonpositive everywhere. In fact $\theta=0$ at $\lambda=0$ since  
the radius (or area) is not changing there. At the world point  
 $C$, it becomes  
negative  and keeps decreasing until reaching negative  
infinity at the caustic 
of $r=0$. ($\theta' \le 0$ all the way once the weak energy condition is  
satisfied.) The initial entropy condition is  
satisfied because $\theta(0)=s(0)=0$. Using the holography bound, we conclude 
that 
$$ 
S_S \ <\  S_B= {1\over 4G} A_B  
$$ 
where $S_S$ is the entropy flux of the star through the horizon and 
$S_B$ ($A_B$) is the entropy (the horizon area) of the final Schwarzschild 
black hole. The equality is not possible because $\theta$ is nonzero 
somewhere.  The interpretation of the result here 
 is that the entropy of star used for the black hole formation  
is less than the final (gravitational) entropy of the  black  
hole.  In other words there is a positive amount of entropy  production 
 along the gravitational collapse.

\section{Holography with timelike sheets} 
 
In the discussion of the holography with lightsheets,  
the Raychaudhuri equation plays an important role. 
The Raychaudhuri equation is, however, not just for the bundle of  
lightlike geodesics but also for  timelike  
geodesics. In this respect here we discuss whether one may 
extend the formulation of holography to the case of timelike 
bulk hypersurface consisting of  timelike  
geodesics.

Let us consider bundle of 
timelike geodesics 
orthogonally  
generated from a codimension two spacelike  
hypersurface $B$ and terminating on $B'$ 
before reaching caustics  or singularities. 
We shall denote the timelike geodesic by 
 $\xi^a=(\partial/\partial \tau)^a$ (with $\xi_a\xi^a=-1$)  
where $\tau$ is the proper time 
along the geodesic.  
We introduce the entropy flux density 
$s$ by 
\begin{equation} 
s= n_a s^a\, \,, 
\label{ttwo} 
\end{equation} 
where $n_a$ is the  normal vector to the timelike  
bulk hypersurface.  
The entropy on the timelike bulk hypersurface 
is then  
\begin{equation} 
S_{T(B-B')}= \int dy^{D-2}\sqrt{h(y)}\int^{\tau'}_0  
d\tau\, {\cal B}\, \, s\,, 
\label{tthree} 
\end{equation} 
where $\vec{y}$ is the coordinate for the boundary 
and 
 ${\cal B}$   
represents the growth of the unit area along $\tau$  
given by 
\begin{equation} 
{\cal B}= exp\left[{\int^\tau_0 d\tilde\tau \,\,\, 
\theta_A(\tilde\tau)}\right] \,, 
\label{tfive} 
\end{equation} 
with the area expansion  
$\theta_A$.  
Note that there are two possible choices for the normal vector  
field $n_a$.  We shall always take the one that makes 
$S_{T(B-B')}$ nonnegative. Infinitesimally away from the timelike  
hypersurface to the normal direction, we extend the geodesics such that  
geodesics are still hypersurface orthogonal. The hypersurface  
orthogonality condition is rather strict and fix the extension 
in the infinitesimal region.  Note that, in the timelike case,
 $\theta = \nabla_a k^a$  
describes not the area expansion but the volume expansion rate. 
We introduce 
the coordinate vector $m^a =(\partial /\partial s)^a$  
whose direction agrees with $n^a$ at the point on the timelike surface
where 
we evaluate  
the expansion rate. By straightforward computation\cite{BKY},  
one may show that 
\begin{equation} 
 \theta=\theta_A+ \theta_m\,, 
\label{va} 
\end{equation} 
where  $\theta_m$ denotes the expansion of the normal vector,
${d\over d\tau} \ln |m|$. 
The timelike bulk 
hypersurface is further restricted by the initial condition 
\begin{equation} 
  s(0)\  \ \le \   -{1\over 4G}\theta(0)\,, 
\label{initial} 
\end{equation} 
and the condition on the expansion of the vector $m^a$ 
\begin{equation}
 \theta_m(\tau) \ge 0 
\,.
\label{normal} 
\end{equation} 
We shall call the resulting hypersurface  
as ``timelike sheet''. At first sight,  the latter condition appears  
too restrictive. However, 
we note that the effect of tidal force in general makes size of body grow
when it is attracted toward a distribution of mass.
This implies that $\theta_m$ is nonnegative rather generically for 
the geodesics of massive objects attracted toward the mass distribution.
We shall give  detailed examples of this elsewhere\cite{BKY}. 
 
Now the proposal is simple.  
The entropy passing through the timelike sheet should be constrained by 
 the difference of the boundary area divided  
by $4G$; 
\begin{equation} 
S_{T(B-B')}\ \le\  
 {1\over 4G}(A-A')\, \,, 
\label{tone} 
\end{equation} 
where $T(B-B')$ denotes the timelike sheet.

For the proof (valid in the classical regime), we 
 shall use the local entropy condition 
$$  
{\rm iii.}\ \  \dot{s}\  \le\     
2\pi \left(T_{ab} \xi^a \xi^b+ {1\over D-2}T\right)   
$$ 
where dots represent derivatives with respect to $\tau$  
and $T= T^a\!_a$. 
As in the lightlike case, the initial condition (\ref{initial}) 
is not for the initial entropy flux but for the choice of 
the initial boundary surface.

The holography here is much stronger version  
than the previous ones, 
and implies 
the lightlike holography we discussed before. 
Namely if the holography is used for  
the timelike sheet that consists of almost  
lightlike geodesics, 
the lightlike holography follows as the limiting case. 
Therefore the timelike holography needs  
 a stronger requirement  on the local entropy flux. 
  Considering  the almost 
 lightlike geodesics, 
the expansion of the normal vector, $\theta_m$, approaches zero and
the condition (iii) implies the condition 
(i). 
 
The proof of the above in the classical regime is  
straightforward. The Raychaudhuri equation reads 
\begin{equation} 
\dot\theta= -{1\over D-1}\theta^2 - \sigma_{ab}\sigma^{ab}   
+\omega_{ab}\omega^{ab} 
 -8\pi G 
\left(T_{ab} \xi^a \xi^b+ {1\over D-2}T\right)\, \,, 
\label{tseven} 
\end{equation}    
where we used the Einstein equation. Since  
 geodesics  are hypersurface orthogonal, 
the twist $\omega_{ab}$ is zero automatically. 
 
Again we note that the differential  
form  of the holography, 
\begin{equation} 
s   \ \,\, \le\   -{1\over 4 G} \theta \ \,\, \le  
-{1\over 4 G} \theta_A 
\,, 
\label{teight} 
\end{equation} 
implies the   integral version, where the second inequality
follows from the condition (\ref{normal}).
             
 Using (\ref{tseven}) and the condition (iii), we have 
\begin{equation} 
-{1\over 4G} \dot\theta = {1\over 4G(D-1)}\theta^2  
+{1\over 4G} \sigma_{ab}\sigma^{ab}+ 
2\pi \left( T_{ab} k^a k^b +{1\over D-2}T \right) 
\ \ge\    \dot{s}      \,. 
\label{tnine} 
\end{equation} 
Together with the initial condition, one has 
\begin{equation} 
s=s_0+\int_0^\tau d \tilde\tau\,\,\dot{s} (\tilde\tau)\   
\le\  -{1\over 4G} 
\left(\theta_0 + \int_0^\tau\, d\tilde\tau\,\,  
\dot\theta (\tilde\tau)\right)=  -{1\over 4G}\theta  \,. 
\label{tten} 
\end{equation} 
 
Like the case of the lightlike holography,  so called 
spacelike projection theorem  holds if certain conditions are  
met\cite{BO}. 
If the future directed timelike bulk hypersurface is complete 
i.e. $B$ is the only boundary,  
the entropy on the 
spacelike bulk enclosed by the boundary would be limited by the boundary  
area divided by $4G$. This  is a consequence of the  
generalized second law. Therefore if any future directed  
timelike sheet is  
complete,  
the original 't Hooft version of the holographic principle  
follows.

Main difference between the lightlike and the timelike holography 
lies in the form of the relevant energy conditions. 
For the lightlike case, 
the null energy condition, 
$$ 
T_{ab} k^a k^b \ \ge\  0 \ \ \ \ {\rm for\  all\  null\ } k^a\,,   
$$ 
is relevant. 
Let us consider a diagonalizable  
energy-momentum tensor that takes the form 
\begin{equation} 
T_{ab}= \rho\,\, t_a t_b +\sum^{D-1}_{i=1}p_i\,\,  x_a^{(i)} x_b^{(i)}  \,, 
\label{t1} 
\end{equation} 
where $(t_a,\ x_a^{(i)})$ is an orthonormal basis with $t_a$  
timelike and $x_a^{(i)}$ spacelike. 
Then the null energy condition implies 
$$ 
\rho + p_i \ \ge\  0 \ \ {\rm for}\ i= 1, \cdots, D-1\,. 
$$   
On the other hand, the energy condition related to the 
the timelike holography is the strong energy condition 
$$ 
T_{ab} \xi^a \xi^b +{1\over D-2}T \ \,\,\,\ge\  \,\,\,  
0 \ \ \ \ {\rm for\  all\   
timelike\ } \xi^a\,.   
$$ 
For the above form of the energy momentum tensor, 
the strong energy condition implies 
$$ 
(D-3)\rho +\sum_i p_i\ \ge\  0\ \ {\rm and}\ \  \rho + p_i \ \ge\  0 \ \  
{\rm for}\ i= 1, \cdots, D-1\ \,. 
$$   
It is generically believed that, for any classically reasonable matter, 
the energy density $T_{ab}\xi^a\xi^b \ge 0$ for any timelike vector  
$\xi^a$. This energy condition is referred as weak energy  
condition. 
In term of the energy density and momentum, the weak energy condition  
implies 
$$ 
\rho \ \ge\  0\ \ {\rm and}\ \  \rho + p_i\  \ge\  0 \ \  
{\rm for}\ i= 1, \cdots, D-1\ \,. 
$$   
The strong energy condition is violated  for the energy momentum  
tensor of positive cosmological  
constant. Due to the relation $\rho= -p_i\  >\  0$, 
one of the strong energy condition is violated by 
$(D-3)\rho +\sum_i p_i= -2 \rho \ < \ 0$.

The null energy  or the strong energy conditions are not  
necessary conditions for the lightlike or the timelike  
holography because 
the derivative of entropy flux density may be positive or  
negative.  

\begin{figure}[ht!] 
\centering \epsfysize=9cm 
\includegraphics[scale=0.8]{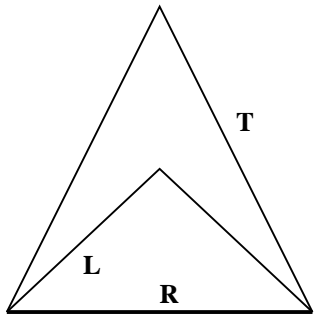} 
\caption{\small Here we consider the case where the 
 timelike-sheet is complete 
such that the spacelike projection theorem is applicable. The solid line  
at the bottom represents the spacelike region enclosed by  
some closed boundary. $L$ and $T$ are respectively  for the lightsheet  
and the timelike-sheet. 
}  
\end{figure}

Finally let us illustrate one specific example  
in which 
 the timelike holography leads to a stronger  
statement than the lightlike holography. We take the case where 
the future-directed timelike sheet is complete before reaching singularity with the  
initial conditions on the expansion respected. (See Fig. 2.)  
Using the lightlike and the timelike 
holography, one has $S_L\  \le\ {1\over 4G} A$ and  
$S_T\  \le\  {1\over 4G}A$ with $A$ being 
the boundary area. 
In addition, by the  
generalized second law,   $S_L\  \le\  S_T$. Thus, one gets 
$$ 
S_L\  \le\ S_T\  \le\ {1\over 4G} A\,. 
$$ 
This is clearly a stronger statement than the one from the  
lightlike holography 
i.e. $S_L\  \le\   {1\over 4G} 
A$.

\section{Discussions} 
 
In this note, we propose a new version of holographic principle 
which compares the entropy passing through 
a bulk hypersurface consisting of  timelike geodesics  
generated from a boundary surface to the boundary area divided 
by $4G$. We  discuss the corresponding  
entropy condition closely related to the Bekenstein bound as a  
generalization  
of the lightlike case. Based on the condition, we prove our  
proposal 
in the classical regime. 
It may be quite interesting  
to derive our timelike holography from the lightlike holography of  
one higher dimensions via the dimensional reduction with a nonvanishing  
Kaluza-Klein momentum.

Our proposal is stronger than the lightlike case because 
the lightlike case can be obtained by a limiting 
procedure in which the velocity field along the timelike  
geodesics  
 approaches  the light  
velocity arbitrarily closely.  
 
One may wonder whether 
further extension of the holography  
to the one based on 
  spacelike geodesics is   
possible or not.  However the spacelike version, if possible, 
will lead to the 't Hooft version of  
holography for general backgrounds,  
for which we already gave 
the flat FRW universe 
as a counterexample.

There are number of points to be studied further.  
Here we do not 
discuss about  the possible quantum modification of the timelike  
holographic principle. In the lightlike case, an attempt is made in  
Ref. \cite{ST}.  
The Raychaudhuri equation also plays an  
essential role for the proof of the singularity theorem.   
As we discussed, the presence of singularity limits  lightsheets  
and  timelike sheets. Singularities seem to 
 have some connections with the entropy of the system.  
In presence of the cosmological singularities  
like the big bang, the meaning of the generalized second law 
is not quite clear. Namely the entropy of certain region 
may not have any operational meaning because the part of the region may lie  
outside of the observer's particle horizon.  Moreover, 
 the notion of time loses its meaning at the big bang or  
other singularities. 
Can we talk about then the fate of 
degrees of freedom  or entropy there?  
The holographic principle may shed some light 
on how to tackle the singularities.


%
%

\section*{Acknowledgments} 
 We are grateful to Raphael Bousso and Naoki Sasakura  
for useful discussions and conversations. 
This work is supported in part by KOSEF 1998      
Interdisciplinary Research Grant 98-07-02-07-01-5.

\end{document}